\documentclass[prb,twocolumn]{revtex4-1}

\usepackage{amsmath}
\usepackage{amsfonts}
\usepackage{amssymb}
\usepackage{graphicx}
\usepackage{color}
\usepackage{bbold}
\usepackage{verbatim}

\begin{document}

\title{Tunneling Magneto-Thermopower in Magnetic Tunnel Junctions}

\author{Carlos L\'opez-Mon\'is, Alex Matos-Abiague and Jaroslav Fabian}

\affiliation{Institute for Theoretical Physics, University of Regensburg, 93040 Regensburg, Germany}

\date{\today}


\begin{abstract}
Thermally induced spin-dependent transport across magnetic tunnel junctions is theoretically investigated. 
We analyze the thermal analog of Slonczewski's model (as well as its limiting case---Julliere's model) of tunneling magnetoresistance and obtain 
analytical expressions for the junction thermopower and the tunneling magneto-thermopower (TMT). The analytical 
model is tested numerically for the special case of an Al$_2$O$_3$-based MTJ, for which we analyze the dependence of the thermopower and TMT on the relative magnetization orientations, as well as on the barrier height and thickness.
We show that at a certain barrier height TMT vanishes, separating the region of positive and negative TMT. As its electrical prototype, this thermal spin transport model should serve as a phenomenological benchmark for analyzing experimental and first-principles calculations of thermopower in magnetic tunnel junctions. The analytical expressions can be used
as a first estimate of the magneto-thermopower of the junctions using {\it ab initio} band structure data of the junction 
ferromagnets.  
\end{abstract}


\maketitle



\section{Introduction}

Traditionally, electric fields have been used as the main force to induce and explore spin-dependent transport in solid-state systems. This has lead to the fundamental and technologically profitable field of spintronics.~\cite{Zutic_RMP_2004} Nevertheless, in the last years we have witnessed how research on thermally driven spin-dependent transport has flourished, bringing forth a new field known as spin caloritronics, which merges spintronics with classical thermoelectricity.~\cite{Bauer_NatMat_2012} The former deals with the interplay between the charge and spin degrees of freedom of carriers, while the latter with the generation of voltages induced by temperature gradients (and vice-versa).~\cite{Barnard} Therefore, spin caloritronics addresses the interaction between spins and heat curents on the transport properties of a system. Although nominally it is a new field, already in the eighties Johnson and Silbsee performed thermodynamic studies on spin-injection across ferromagnetic-nonmagnetic interfaces.~\cite{Johnson_PRB_1987,Johnson_JS_2003} And more recently, Gravier {\it et al.} have measured spin-dependent heat transport across Co/Cu multilayers.~\cite{Gravier_PRB_2006,Gravier_PRB_2006-2} However, the field came to prominence with the discovery of the so called spin-Seebeck effect.~\cite{Uchida_Nature_2008,Slachter_NatPhy_2010,Uchida_NatMat_2010,Jaworski_NatMat_2010,Huang_PRL_2013}

In this framework, spin-valves have proved to be excellent systems for probing spin caloritronic phenomena.~\cite{Breton_Nature_2011,Walter_NatMat_2011,Liebing_PRL_2011,Zhang_PRL_2012,Lin_NatCom_2012,Dejene_PRB_2012,Wang_PRB_2001,McCann_PRB_2002,Czerner_PRB_2011,Jansen_PRB_2012,Hatami_PRB_2009,Scharf_PRB_2012} An ordinary spin-valve is a heterostructure composed of two ferromagnetic materials separated by a mesoscopic nonmagnetic layer. Varying the relative orientation of the magnetizations of the ferromagnets allows to study the spin-dependent properties of the system. When the nonmagnetic material is an insulator or a semiconductor, spin-valves are usually referred to as a magnetic tunnel junctions (MTJ). In order to explore the spin-dependent thermoelectric properties of a spin-valve, a temperature gradient is applied across the system (see Fig.~\ref{fig-MTJ}), which induces charge, spin and/or heat currents ---or voltages. 

In classical thermoelectricity, a material is characterized by its thermal conductivity, Peltier coefficient and thermopower. In the following, we shall focus exclusively on the latter property. The thermopower ---also known as Seebeck coefficient--- measures the magnitude of an induced thermoelectric voltage in response to a temperature gradient across the material. When the spin degree of freedom is taken into account, in addition to the charge voltage, a spin accumulation might also be induced in the system. Analogously, the spin-Seebeck coefficient measures the magnitude of a spin accumulation induced by the temperature gradient.~\cite{Uchida_Nature_2008,Slachter_NatPhy_2010,Uchida_NatMat_2010,Jaworski_NatMat_2010,Huang_PRL_2013} Furthermore, when dealing with spin-valves, it is commonly studied the dependence of the thermopower on the relative magnetization orientations, which has been dubbed as the magneto-Seebeck effect.~\cite{Walter_NatMat_2011,Czerner_PRB_2011} Similar to the tunneling magnetoresistance, a Tunneling Magneto-Thermopower (TMT) can also be defined.~\cite{Walter_NatMat_2011}

Experimentally, Seebeck spin tunneling has been observed in ferromagnet/insulator/silicon tunnel junctions,~\cite{Breton_Nature_2011} realizing thermal spin-injection into semiconductors. The magneto-Seebeck effect was measured in MgO-based MTJs under heat gradients created optically,~\cite{Walter_NatMat_2011} electrically~\cite{Liebing_PRL_2011,Liebing_APL_2013} and even without an external heating source by using the heat dissipation of the tunneling current.~\cite{Zhang_PRL_2012} Giant thermoelectric effects have been observed in Al$_2$O$_3$-based MTJs.~\cite{Lin_NatCom_2012} Besides MTJs, all metallic spin-valves have also been probed for spin-dependent thermal properties.~\cite{Dejene_PRB_2012}

From the theoretical point of view, spin-dependent thermal transport has been studied in ferromagnet/insulator/ferrmagnet MTJs,~\cite{Wang_PRB_2001}. In particular, a giant magneto-thermopower effect has been predicted by magnon-assisted thermal transport.~\cite{McCann_PRB_2002} {\it Ab initio} calculations have been performed for MgO-based MTJs, where the TMT dependence on temperature was computed,~\cite{Czerner_PRB_2011,Czerner_JAP_2012,Heiliger_PRB_2013} and which has been used to understand previously mentioned experiments.~\cite{Walter_NatMat_2011} Similar calculations have been performed for GaAs-based MTJs.~\cite{Jia_AIP_2012} Also inspired by an aforementioned experiment,~\cite{Breton_Nature_2011} theoretical descriptions of thermal spin transport for ferromagnet/insulator/semiconductor MTJs have also been developed.~\cite{Jansen_PRB_2012,Vera-Marun_arXiv_2013} Finally, theory regarding all metallic junctions has as well been recently studied.~\cite{Hatami_PRB_2009,Scharf_PRB_2012,Chen_EPJB_2012,Takezoe_PRB_2010}

Despite all this theoretical effort, to our knowledge, a more elaborated analytical description of the magneto-Seebeck effect in MTJs is still lacking.~\footnote{Except in the case of W. Lin {\it et al.}~\cite{Lin_NatCom_2012}, were they used Julliere's model~\cite{Julliere_PLA_1975} to provide some insight to their experimental data.
}
On the one hand, Slonczewski~\cite{Slonczewski_PRB_1989} developed an analytical model for describing the tunneling magnetoresistance in ferromagnet/insulator/ferromagnet MTJs. On the other, Mott derived a relation between the thermopower and the energy derivative of the conductance for low temperatures.~\cite{Mott,Cutler_PR_1969} In this paper, we derive analytical expressions for both the thermopower and the TMT by combining Slonczewski's model with Mott's relation. We also study the limiting case---Julliere"s model, which provides robust albeit highly simplified
expression for the junction magneto-thermopower. We believe the formulas obtained here should support both {\it ab initio} calculations and experiments.

The paper is organized as follows: the definitions of the thermopower and the TMT are given in Sec.~\ref{TMT-def}, and the transmission probability of the MTJ is computed in Sec.~\ref{model-sec}. In Sec.~\ref{analytics-sec} we derive the analytical formulas for the thermopower and the TMT, and results for the special case of a Al$_2$O$_3$-based MTJ are presented in Sec.~\ref{results-sec}. Finally, a summary is given in Sec.~\ref{summary}.


\section{Theory}

\subsection{Tunneling Magneto-Thermopower} \label{TMT-def}

In general, the current $I$ across a MTJ induced by a thermal gradient $\nabla T$ is given by~\cite{Callen}
\begin{equation} \label{Ic}
  I = -G S \nabla T,
\end{equation}
where $G$ is the conductance and $S$ the thermopower or Seebeck coefficient. This tunneling current can be computed through~\cite{ACTA}
\begin{equation} \label{Isint}
  I = \frac{1}{e} \int g(E) [f_L(E) - f_R(E)] dE
\end{equation}
where $f_L(E)/f_R(E)$ is the Fermi-Dirac distribution of the left/right electrode,
\begin{equation} \label{GE}
  g(E) = \frac{e^2}{h} \frac{1}{(2 \pi)^2} \int T(E,{\bf k}_{\|}) d^2{\bf k}_{\|},
\end{equation}
and $T(E,{\bf k}_{\|})$ is the transmission probability associated with an electron with energy $E$ and transverse $k$-vector ${\bf k}_{\|}$.

Therefore, combining Eqs.~\eqref{Ic} and~\eqref{Isint} in linear response regime, the thermopower $S$ and the conductance $G$ are given by the following integrals:
\begin{subequations}
\begin{eqnarray}
  S & = & -\frac{1}{G} \int g(E) \left( -\frac{\partial f_0}{\partial E} \right) \left( \frac{E - \mu_0}{eT_0} \right) dE, \label{S-int} \\
  G & = & \int g(E) \left( -\frac{\partial f_0}{\partial E} \right) dE, \label{G-int}
\end{eqnarray}
\end{subequations}
respectively, where $\mu_0$ and $T_0$ are the chemical potential and the temperature of the electrodes in equilibrium, respectively. Performing the change of variable $\epsilon = E - \mu_0$ in the integral in Eq.~\eqref{S-int}, it is straightforward to show that for the thermopower $S$ to be finite, $g(\epsilon)$ must not be an even function, i.e., $g(\epsilon) \ne g(-\epsilon)$.

\begin{figure}
  \includegraphics[scale=0.35]{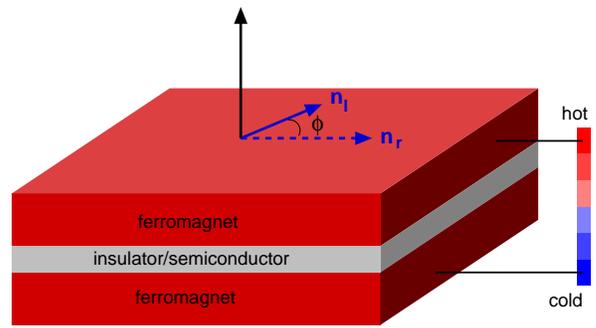}
  \caption{(Color online.) Scheme of a three layer magnetic tunnel junction. A thermally induced current tunnels across the insulator/semiconductor tunneling barrier from one ferromagnet electrode into the other. The thermopower and TMT depend on the relative orientation of the magnetization of the left electrode ${\bf n}_l$ (solid arrow) with respect to the magnetization of the right electrode ${\bf n}_r$ (dashed arrow).}
  \label{fig-MTJ}
\end{figure}

The Tunneling Magneto-Thermopower, which measures the dependence of the thermopower $S$ on the relative in-plane magnetization orientations, is defined as
\begin{equation} \label{TMT}
  {\rm TMT}(\phi) = \frac{S(0) - S(\phi)}{S(\phi)},
\end{equation}
where $\phi$ is the angle spanned between the magnetization vectors of the ferromagnetic layers (see Fig.~\ref{fig-MTJ}).


\subsection{Computation of the transmission probability} \label{model-sec}

In order to study the thermopower, the transmission probability  must be computed. For this purpose we use Slonczewski's model,~\cite{Slonczewski_PRB_1989,ACTA} which describes the ferromagnetic electrodes using the Stoner model,~\cite{Stoner_1938} and the tunneling barrier by means of a rectangular potential. This model also assumes that the energy and transverse modes of the electrons are conserved. The advantage that this description has is that it allows to compute analytically the transmission probability of the MTJ.

Thereupon, the Hamiltonian we use for describing the MTJ is
\begin{equation} \label{Unpert_ham} 
  \mathcal{H} = \left( -\frac{\hbar^2 \nabla^2}{2 m_i^*} + V_i \right) {\cal I} - \frac{\Delta_i}{2} \, \mathbf{n}_i \cdot \boldsymbol{\sigma},
\end{equation}
where the subscript $i$ describes the left ($l$), central ($c$) and right ($r$) layers. The first term in Eq.~\eqref{Unpert_ham} corresponds to the kinetic energy operator, where $m_i^*$ is the electron effective mass in the $i$-th layer. The second term is the rectangular potential barrier with $V_l = V_r = 0$. The third term accounts for coupling between the magnetization and the electron spin in the ferromagnetic leads (Stoner model), where $\Delta_i$ and ${\bf n}_i$ correspond to the exchange energy and a unit vector parallel to the magnetization, respectively, of the $i$-th layer. Since the tunneling barrier is not regarded as ferromagnetic $\Delta_c = 0$. The magnetizations are both in-plane, and we take ${\bf n}_l = (1,0,0)$ and $\mathbf{n}_r = (\cos \phi, \sin \phi, 0)$, where $\phi$ is the angle between the magnetization vectors of the ferromagnetic electrodes. Finally, $\mathcal{I}$ and $\boldsymbol{\sigma} = (\sigma_x,\sigma_y,\sigma_z)$ are the unit matrix in spinor space and the Pauli matrices, respectively.\footnote{
The magnetization orientation is controlled via an external magnetic field. However, the Zeeman splittings are negligible compared to the exchange energy in the ferromagnet, and the orbital effects can be safely neglected as long as the magnetization remains in-plane.~\cite{Wimmer_PRB_2009}
}

Since the transverse modes ${\bf k}_{\|}$ of the conduction electrons are conserved during the tunneling process, the calculation of the transmission probability reduces to a one-dimensional problem. The wave functions found when solving the resulting stationary Pauli-Schr\"odinger equation for a spin-$\sigma$ electron incoming from the left are:~\cite{ACTA}
\begin{subequations}
\begin{eqnarray}
\psi_{l\sigma}(z) & = & \frac{1}{\sqrt{k_{\sigma}}} \,  e^{i k_{\sigma}z} \chi_{l\sigma} \nonumber \\ & + & r_{\sigma,\sigma} e^{-ik_{\sigma}z} \chi_{l\sigma} + r_{\bar{\sigma},\sigma} e^{-ik_{\bar{\sigma}}z} \chi_{l\bar{\sigma}}, \label{phi_i} \\
\psi_{c\sigma}(z) & = & \sum_{i = \pm} \left( C_{\sigma,i} e^{qz} + D_{\sigma,i} e^{-qz} \right) \chi_{li},\\
\psi_{r\sigma}(z) & = & t_{\sigma,\sigma} e^{i\kappa_{\sigma}z} \chi_{r\sigma} + t_{\bar{\sigma},\sigma} e^{i\kappa_{\bar{\sigma}}z} \chi_{r\bar{\sigma}}, \label{phi_t} 
\end{eqnarray}
\end{subequations}
where $k_{\sigma} = \sqrt{k_{\sigma0}^2 - k_{\|}^2}$, $q = \sqrt{q_0^2 + k_{\|}^2}$ and $\kappa_{\sigma} = \sqrt{\kappa_{\sigma0}^2 - k_{\|}^2}$, with
\begin{subequations}
\begin{eqnarray}
  k_{\sigma0} & = & \sqrt{\frac{2m_l^*}{\hbar^2} \left( E + \sigma \frac{\Delta_l}{2} \right)}, \label{k-sig-0} \\
  q_0 & = & \sqrt{\frac{2m_c^*}{\hbar^2} (V_c - E)}, \label{q-0} \\
  \kappa_{\sigma0} & = & \sqrt{\frac{2m_r^*}{\hbar^2} \left( E + \sigma \frac{\Delta_r}{2} \right)}, \label{kappa-sig-0}
\end{eqnarray}
\end{subequations}
and
\begin{equation} \label{eigenvectors-electrodes}
  \chi_{l\sigma} = \frac{1}{\sqrt{2}} \left( \begin{array}{c} 1 \\ \sigma \end{array} \right), \quad \chi_{r\sigma} = \frac{1}{\sqrt{2}} \left( \begin{array}{c} 1 \\ \sigma e^{i\phi} \end{array} \right), \\
\end{equation}
where $\sigma = \uparrow (1), \downarrow (-1)$. The coefficient $t_{\sigma, \sigma}$($t_{\bar{\sigma}, \sigma}$) represents the transmission probability amplitude for a tunneling process in which the electron spin is preserved (flipped). Similarly, $r_{\sigma,\sigma}$ and $r_{\bar{\sigma},\sigma}$ are the reflection probability amplitudes. These amplitudes are computed analytically by solving the set of linear equations obtained when imposing the boundary conditions
\begin{subequations}
  \begin{eqnarray}
    \psi_{i\sigma}(z_{ic}) & = & \psi_{c\sigma}(z_{ic}), \\
    \frac{1}{m_{i}^*} \left. \frac{d\psi_{i\sigma}}{dz} \right|_{z=z_{ic}} & = & \frac{1}{m_c^*} \left. \frac{d\psi_{c\sigma}}{dz} \right|_{z=z_{ic}}
  \end{eqnarray}
\end{subequations}
where $i = l,r$ and $z_{ic}$ is the position of the interface between the central and the $i$-th layer. The transmission probability is now computed through
\begin{equation} \label{T-sigma}
  T_{\sigma}(E,{\bf k_{\|}}) = \frac{m_l^*}{m_r^*} \left( \kappa_{\sigma} |t_{\sigma, \sigma}|^2 + \kappa_{\bar{\sigma}} |t_{\bar{\sigma},\sigma}|^2 \right).
\end{equation}
The total transmission probability is $T = T_{\uparrow} + T_{\downarrow}$.

The computation of the transmission amplitudes is, in general, quite cumbersome. However, in the limit $qd \gg 1$, the following simplified analytical expression for the coefficients $t_{\sigma, \sigma}$ and $t_{\bar{\sigma},\sigma}$ is found:~\cite{ACTA}
\begin{eqnarray} \label{tssp}
  t_{\sigma, \sigma'} & \approx & -\frac{2i m_c^* m_r^* q \sqrt{k_{\sigma}}}{(m_l^*q - im_c^*k_{\sigma})(m_r^*q - im_c^*\kappa_{\sigma})} \nonumber \\ & \times & \left( 1 + \sigma \sigma' e^{-i\phi} \right) e^{-qd},
\end{eqnarray}
which is valid to first order in $\exp(-qd)$. Therefore, replacing Eq.~\eqref{tssp} in Eq.~\eqref{T-sigma} one obtains the transmission probability:
\begin{widetext}
  \begin{eqnarray} \label{T-ana}
    T_{\sigma}(E,k_{\|}) & \approx & \frac{8m_l^*m_r^*m_c^{*2} k_{\sigma}(\kappa_{\sigma} + \kappa_{\bar{\sigma}})(m_r^{*2}q^2 + m_c^{*2} \kappa_{\sigma}\kappa_{\bar{\sigma}})}{(m_l^{*2}q^2 + m_c^{*2} k_{\sigma}^2)(m_r^{*2}q^2 + m_c^{*2} \kappa_{\sigma}^2)(m_r^{*2}q^2 + m_c^{*2} \kappa_{\bar{\sigma}}^2)} \nonumber \\
    & \times & \left[ 1 + \frac{(\kappa_{\sigma} - \kappa_{\bar{\sigma}})(m_r^{*2}q^2 - m_c^{*2} \kappa_{\sigma}\kappa_{\bar{\sigma}})}{(\kappa_{\sigma} + \kappa_{\bar{\sigma}})(m_r^{*2}q^2 + m_c^{*2} \kappa_{\sigma}\kappa_{\bar{\sigma}})} \cos \phi \right] e^{-2qd}.
  \end{eqnarray}
\end{widetext}
Notice that $T(E,{\bf k}_{\|}) = T(E,k_{\|})$.


\subsection{Analytical expression for the TMT} \label{analytics-sec}

To compute the thermopower [Eq.~\eqref{S-int}] and the conductance [Eq.~\eqref{G-int}], the transverse modes $k_{\|}$ in Eq.~\eqref{T-ana} need to be integrated out [Eq.~\eqref{GE}], which can only be done numerically. However, for the case of a high potential barrier ---i.e., $k_{\|} \ll q_0$--- the wave vector $q$ can be approximated as $q = q_0[1 + (k_{\|}/\sqrt{2}q_0)^2]$. In such limit, introducing in Eq.~\eqref{GE} the dimensionless variable $\zeta = k_{\|}^2d/q_0$, due to the exponential factor $\exp(-\zeta)$ in the transmission probability [Eq.~\eqref{T-ana}], the main contribution to the integral in Eq.~\eqref{GE} comes from the vicinity of $\zeta \approx 0$. Therefore, under this approximation Eq.~\eqref{GE} becomes:
\begin{equation} \label{GE-ana}
  g(E) \approx \frac{e^2 q_0}{8\pi^2 \hbar d} \, T(E,0).
\end{equation}
Substituting explicitly the expression for $T(E,0)$ [Eq.~\eqref{T-ana}] in Eq.~\eqref{GE-ana}, gives
\begin{equation} \label{II.302}
  g(E) \approx g_0 \left[ 1 + P_{gl}^{\rm eff}P_{gr}^{\rm eff} \cos \phi \right],
\end{equation}
where $g_0 = g_{l0}g_{r0}$, with
\begin{eqnarray} \label{gl0}
  g_{l0} & = & \sqrt{\frac{2e^2q_0e^{-2q_0d}}{\pi hd}} \\ & \times & \left[ \frac{m_l^*m_c^*(k_{\uparrow 0} + k_{\downarrow 0})(m_l^{*2} q_0^2 + m_c^{*2}k_{\uparrow 0}k_{\downarrow 0})}{(m_l^{*2} q_0^2 + m_c^{*2}k_{\uparrow 0}^2)(m_l^{*2} q_0^2 + m_c^{*2}k_{\downarrow 0}^2)} \right], \nonumber
\end{eqnarray}
and
\begin{equation}  \label{Pleff}
  P_{gl}^{\rm eff} = \frac{(k_{\uparrow 0} - k_{\downarrow 0})}{(k_{\uparrow 0} + k_{\downarrow 0})} \frac{(m_l^{*2} q_0^2 - m_c^{*2}k_{\uparrow 0}k_{\downarrow 0})}{(m_l^{*2} q_0^2 + m_c^{*2}k_{\uparrow 0}k_{\downarrow 0})}, 
\end{equation}
is the effective spin polarization of the left electrode. The expression for $g_{r0}$ ($P_{gr}^{\rm eff}$) is found by replacing in Eq.~\eqref{gl0} [Eq.~\eqref{Pleff}] $k_{\sigma 0}$ and $m_l^*$ with $\kappa_{\sigma 0}$ and $m_r^*$, respectively.

In order now to derive analytical expressions for the thermopower and the TMT, we benefit from Mott's relation,~\cite{Cutler_PR_1969,Mott} which states that
\begin{equation} \label{Mott}
  S = -\frac{\pi^2}{3} \frac{k_B^2}{e} \left. \frac{d}{dE} \log g(E) \right|_{E = \mu_0} T_0,
\end{equation}
where $k_B$ is the Boltzmann constant. Equation~\eqref{Mott} allows to compute the thermopower given the energy dependent conductance $g(E)$. Thus, by replacing Eq.~\eqref{II.302} into Eq.~\eqref{Mott} yields the following expression for the thermopower:
\begin{equation}
  S(\phi) \approx \left[ \frac{1 + (1 - \eta) P_{gl}^{\rm eff}(\mu_0)P_{gr}^{\rm eff}(\mu_0) \cos \phi}{1 + P_{gl}^{\rm eff}(\mu_0)P_{gr}^{\rm eff}(\mu_0) \cos \phi} \right] S_0 \label{Mott-approx}
\end{equation}
where
\begin{equation} \label{eta}
  \eta = \frac{\pi^2}{3} \frac{k_B^2}{e} \frac{1}{S_0} \left. \frac{d}{dE} \log \left( P_{gl}^{\rm eff}P_{gr}^{\rm eff} \right) \right|_{E = \mu_0} T_0,
\end{equation}
is a dimensionless quantity and $S_0 \equiv S(\pi/2)$. The explicit form for the parameter $\eta$ is given in Appendix~\ref{Sec-eta}. Finally, the expression for the TMT is found by replacing Eq.~\eqref{Mott-approx} into Eq.~\eqref{TMT}, which reads
\begin{eqnarray} \label{TMT-ana}
  {\rm TMT}(\phi) & \approx & -\frac{P_{gl}^{\rm eff}(\mu_0)P_{gr}^{\rm eff}(\mu_0)}{1 + P_{gl}^{\rm eff}(\mu_0)P_{gr}^{\rm eff}(\mu_0)} \\ & \times & \frac{\eta(1 - \cos \phi)}{1 + (1 - \eta) P_{gl}^{\rm eff}(\mu_0)P_{gr}^{\rm eff}(\mu_0) \cos \phi}. \nonumber
\end{eqnarray}
Equations~\eqref{Mott-approx} and~\eqref{TMT-ana} represent the thermal analogs of Slonczewski's formulas for the conductance and the tunneling magnetoresistance, respectively.~\cite{Slonczewski_PRB_1989,ACTA} 

An interesting limit to study is that in which Julliere's model is valid.~\cite{Julliere_PLA_1975,ACTA} In such case, the effective spin polarization reduces to~\cite{Julliere_PLA_1975,ACTA}
\begin{equation}
  P_{gi}^{\rm eff} = \frac{D_i^{\uparrow} - D_i^{\downarrow}}{D_i^{\uparrow} + D_i^{\downarrow}}, \quad i = l,r,
\end{equation}
where $D_i^{\sigma}$ is the spin-dependent density of states in the $i$th-electrode. Moreover, since $2g_0 = g(\phi = 0) + g(\phi = \pi)$ [see Eq.~\eqref{II.302}] then,~\cite{ACTA}
\begin{equation}
  g_0 \propto \frac{1}{2} \left( D_l^{\uparrow} D_r^{\uparrow} + D_l^{\downarrow} D_r^{\downarrow} + D_l^{\uparrow} D_r^{\downarrow} + D_l^{\downarrow} D_r^{\uparrow} \right),
\end{equation}
which allows also to compute $S_0$ from the density of states 
and, hence, the $\eta$ parameter [Eq.~\eqref{eta}]. Therefore, in the limit where Julliere's model and Mott's law are valid, one can estimate both the thermopower [Eq.~\eqref{Mott-approx}] and the TMT [Eq.~\eqref{TMT-ana}] only through the density of states.


\section{Results} \label{results-sec}

In this section we discuss the results obtained for the special case of a Fe/Al$_2$O$_3$/Fe MTJ. Since the system is symmetric, the effective spin polarization [Eq.~\eqref{Pleff}] is the same in both electrodes, so $P_{gl}^{\rm eff} = P_{gr}^{\rm eff} \equiv P$. In addition, for the set of parameters discussed throughout this section, the relation $P^2 \ll 1$ is satisfied, hence, Eq.~\eqref{Mott-approx} takes the simple form
\begin{equation}
  S(\phi) \approx [1 - \eta P^2 \cos \phi] S_0. \label{Mott-approx-sym}
\end{equation}
This cosine-like behavior found for the angular dependence of thermopower has been observed in Refs.~[\onlinecite{Liebing_PRL_2011},\onlinecite{Zhang_PRL_2012}], where they measure the induced thermovoltage for CoFeB/MgO/CoFeB MTJs. However, {\it ab initio} calculations performed in Ref.~[\onlinecite{Czerner_PRB_2011}] for the same kind of MTJs show a different behavior. Furthermore, replacing Eq.~\eqref{Mott-approx-sym} in the general expression for the TMT [Eq.~\eqref{TMT}] yields
\begin{equation}
  {\rm TMT}(\phi) \approx -\eta P^2 \, (1 - \cos \phi), \label{TMT-ana-sym}
\end{equation}
where we have used that $\eta P^2 \ll 1$.


\begin{figure}
  \includegraphics[width=\columnwidth]{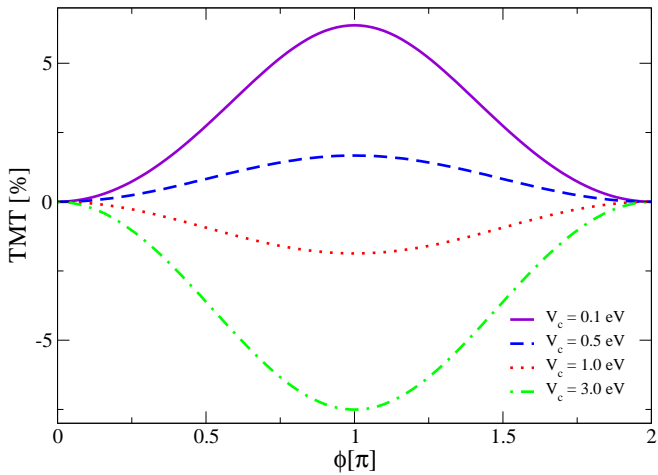}
  \caption{(Color online.) TMT dependence of a Fe/Al$_2$O$_3$/Fe MTJ on the relative orientation of the magnetizations of the ferromagnets, for different values of the tunneling barrier height $V_c$, and thickness $d = 20\,$\AA. The values used for the remaining model parameteres are: $m_l^* = m_r^* = m_0$ and $m_c^* = 0.4m_0$, where $m_0$ is the bare electron mass, and $k_{\uparrow} = 1.09\,$\AA$^{-1}$ and $k_{\downarrow} = 0.42\,$\AA$^{-1}$.
  }
  \label{TMT-phi}
\end{figure}

Figure~\ref{TMT-phi} shows the TMT dependence on the relative magnetization $\phi$ for different values of the barrier height $V_c$. The behavior observed is well described by Eq.~\eqref{TMT-ana-sym}. Notice how the sign of the TMT changes as the value of $V_c$ increases. This is a novel feature compared to the behavior found for the tunneling magnetoresistance, which is always positive.~\cite{ACTA} Let us now discuss more in detail this sign change.


\begin{figure}
  \includegraphics[width=\columnwidth]{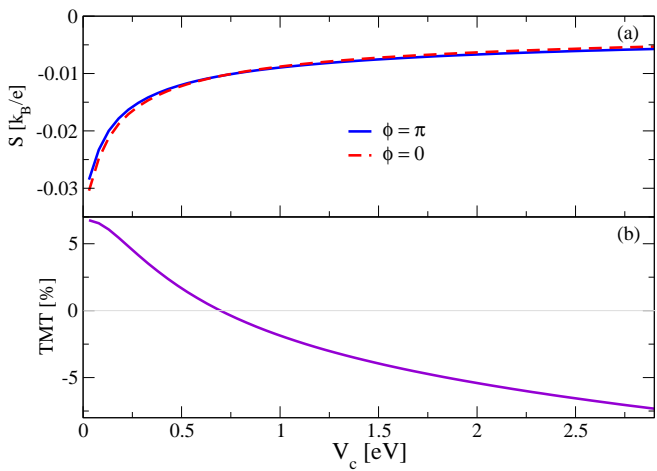}
  \caption{(Color online.) (a) Thermopower of a Fe/Al$_2$O$_3$/Fe MTJ as a function of the barrier height. The blue solid (red dashed) line corresponds to the case where the magnetizations are antiparallel (parallel). (b) TMT$(\pi)$ dependence on the barrier height. Notice that the thermopower for the parallel and antiparallel orientations cross, meaning the TMT becomes zero. Barrier thickness $d = 20\,$\AA. Idem as Fig.~\ref{TMT-phi}.}
  \label{S-TMT-Vc}
\end{figure}

Figure~\ref{S-TMT-Vc}(a) shows the thermopower dependence on the barrier height for the cases when the magnetizations are parallel, $\phi = 0$ (red dashed line), and antiparallel, $\phi = \pi$ (blue solid line). Figure~\ref{S-TMT-Vc}(b) shows the ${\rm TMT}$ dependence on the barrier height when $\phi = \pi$. The behavior found is not as straightforward to understand as in the case of the angular dependence, since the TMT is a complicated function of $V_c$. Figure~\ref{S-TMT-Vc}(b) shows that the TMT changes sign for a specific value of barrier height $V_c^{(0)}$, which corresponds to the crossing between the thermopower for the parallel and the antiparallel cases [see Fig.~\ref{S-TMT-Vc}(a)]. The condition for ${\rm TMT}(\phi) = 0$ [Eq.~\eqref{TMT-ana}] is satisfied when $P_{gi}^{\rm eff}(\mu_0) = 0$, which according to Eq.~\eqref{Pleff} occurs when
\begin{equation}
  V_{c}^{(0)} = \frac{\hbar^2 k_{\uparrow}k_{\downarrow}}{2(m_0^{2}/m_c^*)} + \mu_0,
\end{equation}
where $k_{\sigma} = k_{\sigma 0} (\mu_0) = \kappa_{\sigma 0} (\mu_0)$ and $m_l^* = m_r^* = m_0$, where $m_0$ is the bare electron mass.
For this same value the tunneling magnetoresistance also becomes zero,~\cite{ACTA} although it remains positive, as mentioned before, the reason being that the sign change of the TMT is not related to the effective spin polarization ---since for a symmetric MTJ the TMT is a function only of $P^2$ [Eq.~\eqref{TMT-ana-sym}]--- but to a sign change in the parameter $\eta$.


\begin{figure}
  \includegraphics[width=\columnwidth]{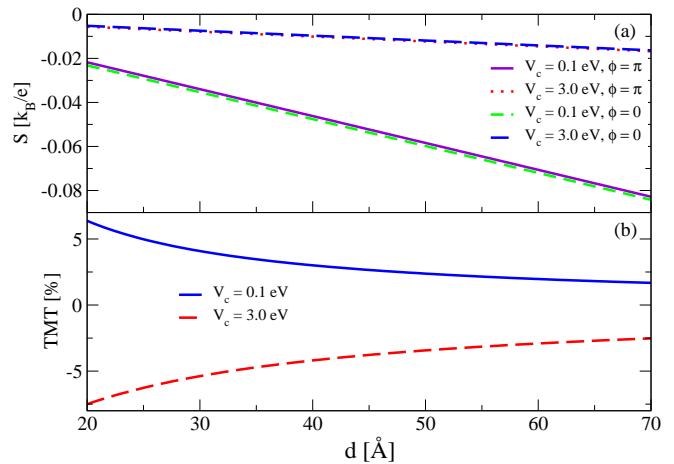}
  \caption{(Color online.) (a) Thermopower of a Fe/Al$_2$O$_3$/Fe MTJ as a function of the barrier thickness. (b) TMT$(\pi)$ dependence on the barrier thickness. Idem as Fig.~\ref{TMT-phi}.
}
  \label{S-TMT-d}
\end{figure}

Figure~\ref{S-TMT-d}(a) [(b)] shows the thermopower (TMT) dependence on the barrier thickness $d$ when the magnetizations are parallel and antiparallel. The barrier thickness only enters in the denominator of $\eta$ (see Appendix~\ref{Sec-eta}). Therefore, as $d$ increases $\eta \to 0$ and, hence, $S \to S_0$ [Eq.~\eqref{Mott-approx}], which increases linearly with $d$, and the ${\rm TMT} \to 0$ [Eq.~\eqref{TMT-ana}], as seen in Figs.~\ref{S-TMT-d}(a) and (b), respectively. Notice that while the thermopower increases with the barrier thickness, the TMT decreases. This behavior agrees with the low temperature trend found in the {\it ab initio} calculations in Ref.~[\onlinecite{Czerner_JAP_2012}] for a CoFeB/MgO/CoFeB MTJ, where the dependence on the thickness of the barrier was studied by varying the number of MgO monolayers.


All figures discussed in this section were produced using the analytical Eqs.~\eqref{Mott-approx} and~\eqref{TMT-ana} for the thermopower and the TMT, respectively. In addition, they have been crosschecked by numerically integrating Eqs.~\eqref{S-int} and~\eqref{G-int} with $g(E)$ given in Eq.~\eqref{II.302}. The temperature for the former calculation was of $4.2\,$K. The agreement found between both computations was extremely good.


Finally, for a symmetric MTJ the effective spin polarization $P$ can be extracted from tunneling magnetoresistance experiments, by measuring the parallel and antiparallel conductances $G_P$ and $G_{AP}$, respectively, through the equation:~\cite{ACTA}
\begin{equation} \label{P-exp}
  P = \pm \sqrt{\frac{G_P - G_{AP}}{G_P + G_{AP}}}.
\end{equation}
Therefore, by measuring the parallel, $S_P$, and antiparallel, $S_{AP}$, values of the thermopower, it is possible to experimentally estimate the parameter $\eta$ with the equation
\begin{equation}
  \eta = -\frac{2G_PG_{AP}}{G_PS_P + G_{AP}S_{AP}}\frac{S_P - S_{AP}}{G_P - G_{AP}},
\end{equation}
found when replacing Eq.~\eqref{P-exp} in Eq.~\eqref{TMT-ana}.


\section{Summary} \label{summary}

We have studied thermal spin transport in magnetic tunnel junctions using the thermal analog of Slonczewski's model
of tunneling. We have derived analytical expressions for the thermopower and the Tunneling Magneto-Thermopower
of magnetic junctions in both Slonczewski's approximation and in the limit of Julliere's model. We show that TMT 
can be both positive and negative, depending on the barrier properties, crossing through zero at a certain barrier
height. Our expressions could be used in combination with first principles band structure parameters of the bulk
ferromagnetic materials forming the junction to estimate the spin thermal transport characteristics. Furthermore, they 
can serve as a phenomenological description of experiments on junction magneto-thermopower, in most cases with 
a single fitting parameter ($\eta$).


\begin{acknowledgments}
The authors are grateful for the financial support offered by the Deutsche Forshungsgemeinschaft (DFG) via the Priority Program ``Spin Caloric Transport'' (SPP 1538).
\end{acknowledgments}


\appendix

\section{The $\eta$ parameter} \label{Sec-eta}

\begin{widetext}

The expression found for $\eta$ is:
\begin{equation}
  \eta = \sum_{i = l,r} \frac{2m_i^*}{\hbar^2 k_{i\uparrow}k_{i\downarrow}} \left( 1 + \frac{m_i^{*2} m_c^{*2}q_F^2 \left( k_{i\downarrow}^2 + k_{i\uparrow}^2 \right) + 2m_i^* m_c^{*3} k_{i\uparrow}^2k_{i\downarrow}^2}{m_i^{*4} q_F^4 - m_c^{*4}k_{i\uparrow}^2k_{i\downarrow}^2} \right) \left( \left. \frac{d}{dE} \log g_0(E) \right|_{E = \mu_0} \right)^{-1},
\end{equation}
where
\begin{eqnarray}
  \left. \frac{d}{dE} \log g_0(E) \right|_{E = \mu_0} & = & \frac{2m_c^*}{\hbar^2q_F^2} \left(q_Fd - \frac{1}{2} \right) + \frac{1}{\hbar^2} \sum_{i = l,r} \Bigg[ \frac{m_i^*}{k_{i\uparrow}k_{i\downarrow}} \left(1 + \frac{m_c^{*2}(k_{i\uparrow}^2 + k_{i\downarrow}^2) - 2m_i^*m_c^*k_{i\uparrow}k_{i\downarrow}}{m_i^{*2} q_F^2 + m_c^{*2}k_{i\uparrow}k_{i\downarrow}} \right) \nonumber \\ 
& - & \left( \frac{2m_i^{*2}q_F^2 + m_c^{*2}(k_{i\uparrow}^2 + k_{i\downarrow}^2)}{(m_i^{*2}q_F^2 + m_c^{*2}k_{i\uparrow}^2)(m_i^{*2}q_F^2 + m_c^{*2}k_{i\downarrow}^2)} \right) 2m_i^*m_c^*(m_c^* - m_i^*) \Bigg],
\end{eqnarray}
and $k_{l\sigma} = k_{\sigma 0} (\mu_0)$, $q_F = q_0 (\mu_0)$ and $k_{r\sigma} = \kappa_{\sigma 0} (\mu_0)$. Since the effective masses and Fermi wave-vectors can be computed through {\it ab initio} calculations, it is also possible to estimate the parameter $\eta$.

\end{widetext}

\bibliography{Seebeck}{}

\end{document}